\documentclass[showpacs,preprintnumbers,amsmath,amssymb]{revtex4}
\newtheorem{theorem}{Theorem}
\newtheorem{definition}{Definition}
\newtheorem{lemma}{Lemma}
\newtheorem{corollary}{Corollary}

\usepackage{graphicx}
\usepackage{dcolumn}
\usepackage{bm}

\begin{document}
\title{
Quantum Kolmogorov Complexity and Bounded Quantum Memory
}

\author{Takayuki Miyadera}
\affiliation{%
Research Center for Information Security (RCIS), \\
National Institute of Advanced Industrial
Science and Technology (AIST). \\
Daibiru building 1003,
Sotokanda, Chiyoda-ku, Tokyo, 101-0021, Japan.
\\
(E-mail: miyadera-takayuki@aist.go.jp)
}%


\date{\today}

\begin{abstract}
In this study, the effect of bounded quantum memory 
in a primitive information protocol 
has been examined 
using the quantum Kolmogorov complexity as a measure of information. 
We employed a 
toy two-party protocol in which Bob by using 
a bounded quantum memory and an unbounded classical memory 
estimates a message that was encoded in qubits 
by Alice in one of the bases $X$ or $Z$. 
Our theorem gave a nontrivial effect of the memory boundedness. 
In addition, a generalization 
of the uncertainty principle in the presence of quantum memory
has been obtained. 
\end{abstract}
\pacs{03.67.Lx, 03.65.Ta, 03.67.Dd}
\maketitle
\section{Introduction}
Various ideas have been put forth and 
experiments have been conducted to 
construct a large and stable quantum memory. 
This is definitely the most 
important factor for achieving quantum information processing 
devices. 
However, despite the efforts expended thus far, it is 
 difficult to imagine that 
any ultimate method will be proposed that enables the 
construction of an arbitrarily large quantum memory. 
\par
This pessimistic view is not as disappointing 
as it may seem. 
Let us consider a two-party cryptographic protocol 
called oblivious transfer \cite{Rabin}, in which 
Alice sends a bit to Bob 
by executing a protocol 
in such a manner that 
Bob receives it correctly with probability $1/2$ 
and nothing otherwise, but 
Alice cannot learn whether her bit was received. 
While the oblivious transfer
could be a strong cryptographic primitive 
when realized, 
its secure realization 
without considering any assumptions 
is impossible 
even if one uses quantum communication
\cite{LoOblivious,MayersOblivious}.  
Several studies have been conducted to 
investigate physical assumptions yielding 
the secure implementation of this protocol 
using classical communication. 
The first assumption is the boundedness of computation power, 
which is the most traditional approach based on 
a mathematically unproven 
computational complexity problem \cite{Rabin}. 
This approach implicitly contains 
a physical assumption: it assumes the 
processing speed of existing computers. 
The second assumption involves 
the imperfection of a communication channel. 
It has been demonstrated in various noise models that 
secure oblivious transfer is achievable
\cite{Crepeau88,Damgaard2004}. 
The third assumption, which is related to the present 
study, is the boundedness of memory size. 
It has been shown that the protocol may be
securely implemented if the size of 
an adversary's (classical)
memory is restricted \cite{Cachin}.  
Although this result is interesting in theory, 
the assumption is not entirely valid because 
it is not too difficult to construct 
large classical memories with 
current technology. 
Recently, important work based on 
bounded quantum memory has been completed. 
In \cite{Damgaard}, 
Damgaard et al. have shown that 
the oblivious transfer becomes information-theoretically 
secure under the assumption that the size of the 
adversary's quantum memory is bounded. 
As mentioned earlier, the assumption of bounded 
quantum memory is more reasonable than 
that of bounded classical memory.
In fact, the latest technology enables the stability of 
only a few 
qubits of memory. 
\par
The
protocol runs as follows. 
Alice encodes a random $N$ bit sequence 
to a quantum state of $N$ qubits in $X$ or $Z$ basis.
She sends the qubits to Bob. 
Bob selects $X$ or $Z$ and measures 
the qubits with respect to the selected basis. 
After Alice announces the basis used
for encoding, Bob knows whether his selection of 
the basis is correct.
The privacy amplification phase 
 causes the players to agree on a bit in the case that the 
basis is correct, otherwise the protocol makes Bob 
completely ignorant of the bit obtained by Alice. 
Alice can send Bob an arbitrary bit by XORing 
with her obtained bit. As intended, this method
 works as an oblivious channel. 
To demonstrate the security, 
assume that one of the players, Bob, is 
dishonest. 
If Bob has a quantum memory of size larger than 
$N$, he may not follow the protocol but 
may keep the quantum state and measure
it with respect to the disclosed basis. 
This procedure provides the full information 
of an encoded sequence to Bob; thus the protocol 
fails. 
Further, imagine a case in which 
dishonest Bob's quantum memory is bounded. 
Since he cannot keep the entire quantum state, 
he has to irreversibly convert a part of the 
quantum state to the classical state by measuring 
a part of qubits. 
This operation destroys the quantum state 
and the information encoded in it. 
It is essential that he is unaware of 
the basis used for encoding when the qubits are passed to him.  
In view of the fact that 
the noncommutativity of $X$ and $Z$ prohibits their 
joint measurement, any measurement inevitably 
induces loss of information. 
For instance, suppose that Bob with $M (<N)$ qubits memory 
employs the following simple strategy. 
When the qubits are sent, 
Bob keeps the first $M$ qubits as they are 
and measures the remaining $N-M$ qubits 
in the $Z$ basis. 
This strategy allows Bob to obtain full information 
if the basis used by Alice for encoding 
was $Z$. Conversely, it does not provide 
full information to Bob if the basis used by 
Alice was $X$, because he
loses the information that was encoded 
in the first $M$ qubits. 
Damgaard et al. \cite{Damgaard} has shown that 
dishonest Bob must have at least $N/2$ 
qubits quantum memory to benefit 
in this protocol. 
That is, the protocol works 
under the assumption of bounded quantum memory 
of size smaller than $N/2$. 
\par
In this study, we inspect a toy two-party protocol
that corresponds to the oblivious transfer protocol 
with dishonest Bob, 
but is not followed by privacy amplification. 
Consider the following problem. 
Assume that Alice encodes a random $N$ bit sequence 
to a quantum state in the $X$ or $Z$ basis. 
How much information can 
Bob, who has a bounded quantum memory and 
an unbounded classical memory, 
retrieve after the basis is announced? 
By extracting part of a protocol in this manner, 
it is hoped that   
the understanding of the power of bounded quantum memory 
can be deepened. 
In addition, it may help improve the existing result, 
as was the case for the quantum key distribution protocol. 
We study the problem by employing the quantum Kolmogorov 
complexity as a measure of information, as 
defined by Vit\'anyi \cite{Vitanyi}.
In contrast to Shannon's theory, 
which treats only information of 
probabilistic sources, 
the Kolmogorov complexity is often called absolute information because it 
can assign information to 
individual objects. 
The key notion is algorithmic randomness. 
The complexity of an object is defined as the shortest 
description length in both classical Kolmogorov 
complexity and 
quantum Kolmogorov complexity given by Vit\'anyi. 
While the quantum Kolmogorov complexity 
is a natural concept for absolute information, 
it has been used in quantum information theory 
to develop only a few applications 
\cite{MiyaQKD,MiyaEnt}
developed in the quantum information theory. 
One of the purposes of this study 
 is to demonstrate the usefulness of the quantum Kolmogorov 
 complexity for analyzing
protocols in this area. 
\par
The next section contains a brief review of the 
quantum Kolmogorov complexity as defined by Vit\'anyi. 
In section \ref{section:main}, we introduce a toy 
two-party protocol and describe our main result based on it. 
The conclusion contains a discussion of the results. 
\section{Quantum Kolmogorov Complexity based on Classical Description}
The classical Kolmogorov complexity of
a binary sequence is defined by 
the length of the shortest program, as proposed by 
Kolmogorov \cite{Kolmogorov} 
and Chaitin \cite{Chaitin} independently, for 
a one-way Turing machine to output the sequence. 
For instance, a binary sequence $``000\ldots00"$ has 
a short description such as ``print $``0"$ $10000$ times," 
which therefore has a small Kolmogorov complexity. 
If a binary sequence has no pattern that yields any 
compressed description, the best way to describe it is 
to write it down naively. Such a sequence is called 
random. That is, a binary sequence is random if and only if 
its Kolmogorov complexity is as large as its own length. 
Although a specification of a Turing machine 
is required to define its value, 
the Kolmogorov complexity does not depend on the choice of a Turing machine 
except for a trivial constant. 
Moreover, the classical Kolmogorov complexity 
and its conditional version have various rational properties, 
as do Shannon entropy and conditional entropy. 
The Kolmogorov complexity plays a central role in 
field of algorithmic information theory whose applications 
extend to vast areas such as the foundation of mathematics, 
computation theory, and physics \cite{LiVitanyi, ChaitinBook}. 
\par
While it seems natural to define the corresponding complexity for 
quantum states, the quantum versions 
\cite{Vitanyi,Svozil,vanDam,Gacs,Muller,MullerPhD}
of the Kolmogorov complexity 
were only recently proposed. 
 Among the several versions of the quantum Kolmogorov 
complexity, we employ the one that was defined by 
Vit\'anyi \cite{Vitanyi}. 
Vit\'anyi's definition based on the classical description length is 
suitable for quantum information-theoretic problems that 
normally handle classical inputs and outputs.   
In order to explain the definition precisely, 
a description of a one-way quantum Turing machine 
is needed. 
A one-way quantum Turing machine consists of 
four tapes and an internal control. 
(See \cite{Vitanyi} for more details.)
Each tape is a one-way infinite qubit (quantum bit) chain
and has a corresponding head on it. 
One of the tapes works as the input tape and 
is read-only from left-to-right.
A program is given on this tape
as an initial condition.
The second tape functions as the work tape.
The work tape is initially set to be $0$ for 
all the cells. The head on this tape can read and write 
a cell, and can move in both directions.
The third tape is called an auxiliary tape.
One can put an additional input on this tape. 
The additional input is written to the left-most qubits and 
can be a quantum state or a classical state. 
This input is needed when one deals with the conditional 
Kolmogorov complexity. 
The fourth tape works as the output tape. 
It is assumed that after halting
 the state over this tape will not be changed.
The internal control is a quantum system 
described by a finite dimensional 
Hilbert space 
that has two special orthogonal vectors
$|q_0\rangle$ (initial state) and $|q_f\rangle$ (halting state). 
After each step one makes a measurement of a coarse grained 
observable on the internal control $\{|q_f\rangle \langle q_f|,
{\bf 1} -|q_f\rangle \langle q_f|\}$ to know if 
the computation halts \cite{Bernstein}. 
A computation halts at time $t$ if and only  
if 
the probability to observe $q_f$ at time $t$ is $1$, 
and at any time $t'<t$ the probability to observe 
$q_f$ is zero (see \cite{Myers, Ozawa, Popescu, Miya} for 
relevant discussions).  
By using this one-way quantum Turing machine,  
Vit\'anyi defined the quantum Kolmogorov complexity as 
the length of the shortest description of a quantum state.
That is, the programs of quantum Turing machine are restricted to 
classical ones, while 
the auxiliary inputs can be quantum states. 
We write $U(p,y)=|x\rangle$ if and only if 
a quantum Turing machine $U$ with 
a classical program $p$ and an auxiliary 
(classical or quantum) input $y$ halts 
and outputs $|x\rangle$. 
The following is the precise description of 
Vit\'anyi's definition.
\begin{definition}\cite{Vitanyi}
The (self-delimiting) quantum Kolmogorov 
complexity of a pure state $|x\rangle$ 
with respect to a one-way quantum Turing machine $U$ with $y$ 
(possibly a quantum state) as conditional input 
given for free is 
\begin{eqnarray*}
K_U(|x\rangle,|\ y)
:=\min_{p,|z\rangle} \{l(p)+ \lceil -\log|\langle z|x\rangle |^2 \rceil :
U(p,y)=|z\rangle\},
\end{eqnarray*}
where $l(p)$ is the length of a classical program $p$, and 
$\lceil a \rceil$ is the smallest integer larger than $a$. 
\end{definition}
The one-wayness of the quantum Turing machine ensures that 
the halting programs compose a prefix free set. 
Because of this, the length $l(p)$ is defined consistently. 
The term $\lceil -\log|\langle z|x\rangle |^2 \rceil$
represents how insufficiently an output $|z\rangle$ approximates 
the desired output $|x\rangle$. 
This additional term has 
a natural interpretation using the Shannon-Fano code. 
Vit\'anyi has shown the following invariance theorem, 
which is very important.  
\begin{theorem}\cite{Vitanyi}
There is a universal quantum Turing 
machine $U$, such that for all machines $Q$, 
there is a constant $c_Q$, such that for all quantum states $|x\rangle$
and all auxiliary inputs $y$ we have:
\begin{eqnarray*}
K_U(|x\rangle  |\ y)
\leq K_Q(|x\rangle |\ y)
+c_Q.
\end{eqnarray*}
\end{theorem}
Thus the value of 
quantum Kolmogorov complexity does not depend on 
the choice of a quantum Turing machine if 
one neglects the unimportant constant term $c_Q$. 
Thanks to this theorem, 
one often writes $K$ instead of 
$K_U$. 
Moreover, the following theorem is crucial for our discussion.
\begin{theorem}\label{th:classical}\cite{Vitanyi}
On classical objects (that is, finite binary strings that are all directly 
computable) the quantum Kolmogorov complexity 
coincides up to a fixed additional constant with the self-delimiting 
Kolmogorov complexity. That is, 
there exists a constant $c$ such that for 
any classical binary sequence $|x\rangle$,
\begin{eqnarray*}
\min_q \{l(q): U(q,y)=|x\rangle \}\geq
 K(|x\rangle|\ y)\geq 
\min_q \{l(q): U(q,y)=|x\rangle\} -c
\end{eqnarray*}
holds.
\end{theorem}
According to this theorem, for classical objects 
it essentially suffices to treat only programs that exactly 
output the object.
\section{Formulation and results}\label{section:main}
In order to discuss the power of the bounded quantum memory, 
we use a toy two-party protocol. 
Suppose that there exist two players
Alice and Bob. 
Alice encodes a message in qubits 
with one of the bases $X$ or $Z$, 
and 
sends them to Bob. 
The precise formulation is as follows. 
Alice chooses probabilistically \cite{prob} an
$N$-bit message $x\in \{0,1\}^N$. 
She also chooses a basis $X$ or $Z$ for its encoding. 
We write the standard basis of 
a qubit as $\{|0\rangle, |1\rangle\}$, 
which are eigenstates of $Z$. 
Its conjugate basis is written as 
$\{|\overline{0}\rangle, |\overline{1}\rangle\}$, 
which are eigenstates of $X$ and are defined as 
$|\overline{0}\rangle :=
\frac{1}{\sqrt{2}}(|0\rangle +|1\rangle)$
and $|\overline{1}\rangle:=
\frac{1}{\sqrt{2}}(|0\rangle -|1\rangle)$. 
She prepares a quantum state 
of $N$ qubits described by a Hilbert space 
${\cal H}_A$ as follows. 
If her choice of basis is $X$,
she encodes her message $x=x_1x_2\cdots x_N \in \{0,1\}^N$
as $|\overline{x}\rangle:=
|\overline{x_1}\rangle \otimes |\overline{x_2}\rangle
\otimes \cdots \otimes |\overline{x_N}\rangle \in {\cal H}_A$. 
Conversely, if her choice of basis is $Z$, 
she prepares $|x\rangle:=
|x_1\rangle \otimes |x_2\rangle \otimes \cdots \otimes |x_N\rangle
\in {\cal H}_A
$. 
We define $X_x:=|\overline{x}\rangle \langle \overline{x}|$ 
for each $x\in \{0,1\}^N$ and $Z_z:=
|z\rangle \langle z|$ for each $z\in \{0,1\}^N$. 
Bob, who has a bounded quantum memory and 
an unbounded classical memory, tries estimating 
the message after the basis is disclosed by Alice.
There may exist some different formulations of the quantum 
bounded memory. For instance, one of the possible definitions 
would be such that Bob has many qubits that 
decohere in an uncontrollable way except for some 
$M$ qubits. 
On the other hand, the following definition employed in 
this paper gives Bob the ability to control the system. 
Bob has an arbitrarily large system and he makes it interact with 
the qubits sent from Alice. 
The whole system that Bob possesses after the interaction 
is divided into two parts as ${\cal H}_m \otimes {\cal K}$. 
The first part is 
the quantum memory that 
consists of $M$ qubits. The second part is called 
an auxiliary part whose size can be arbitrarily large. 
Because Bob cannot keep the quantum coherence of the auxiliary part, 
he makes a measurement of an observable on it before the basis 
is announced by Alice. 
Let us denote the observable by $C=\{C_{\xi}\}$, which forms 
a positive-operator-valued measure (POVM) on ${\cal K}$. 
The measurement result is stored in a classical memory 
whose size can be arbitrarily large. 
For as long as he needs, 
Bob can keep the quantum state on the quantum memory. 
He estimates the message 
encoded by Alice by using the quantum state 
in the quantum memory, the classical data in 
the classical memory and the basis disclosed by Alice. 
The whole process can be written as follows. 
The interaction between the qubits sent by Alice and 
Bob's apparatus is described by a completely positive map
(CP-map), 
\begin{eqnarray*}
\Lambda: \Sigma({\cal H}_A)
\to \Sigma({\cal H}_m \otimes {\cal K}), 
\end{eqnarray*}
where $\Sigma({\cal H})$ 
denotes a set of all 
density operators on ${\cal H}$. 
For instance, if Alice uses the basis $Z$ for encoding 
message $z$, the state after the interaction 
becomes $\Lambda(|z\rangle \langle z|)$. 
Suppose that Bob obtained $\xi$ as an outcome.
The quantum state kept in the quantum memory 
depends on $Z$, $z$, and $\xi$, and is denoted by 
$\rho^Z_{z,\xi} \in \Sigma({\cal H}_m)$, which can be calculated as 
an a-posteriori state \cite{aposteriori}.
We treat the quantum Kolmogorov complexity 
$K(z|\rho^Z_{z,\xi},\xi,Z)$ as a measure to 
characterize Bob's estimation \cite{identify}. 
That is, if Bob can estimate $z$ exactly only from 
what he actually has, $K(z|\rho^Z_{z,\xi},\xi,Z)
=O(1)$ holds. Otherwise, the quantum Kolmogorov complexity 
becomes nontrivial. 
Similarly, when Alice uses the basis $X$ for 
encoding message $x$, and Bob obtains $\xi$ as 
an outcome of the measurement on the auxiliary system, 
we denote the quantum state in the quantum memory by 
$\rho^X_{x,\xi} \in \Sigma({\cal H}_m)$. 
$K(x|\rho^X_{x,\xi},\xi,X)$ characterizes 
Bob's ability to estimate the message $x$ \cite{identify}.
\par
When $M\geq N$ holds, Bob can recover the message perfectly 
by simply keeping the quantum states sent by Alice. 
We study how well Bob can recover the message 
when $M <N$ is satisfied. 
\par
In the following theorem, 
we study asymptotic behavior 
with respect to increasing $N$. For each $N$, the size of 
Bob's quantum memory
$M$, which may depend on $N$, is bounded as $M\leq qN$ for some $q$ 
$(0\leq q\leq 1)$. 
For each $N$, 
$P(x|\xi_N,X)$ denotes a posterior probability 
of a message $x\in \{0,1\}^N$ when Alice 
used the basis $X$ and Bob obtained outcome $\xi_N$. 
Similarly, 
$P(z|\xi_N,Z)$ denotes 
a posterior probability of a message 
$z\in \{0,1\}^N$ when Alice used the 
basis $Z$ and 
Bob obtained $\xi_N$. 
\begin{theorem}\label{assym}
Let us consider a family of protocols indexed by the number of qubits $N$. 
Assume that for each $N$ 
the size of Bob's quantum memory $M$, which may 
depend on $N$, is bounded as $M\leq qN$ for some $q$ $(0\leq q\leq 1)$.
For any $p_X$ and $p_Z$ satisfying
$q+ p_X +p_Z <1,$ 
there exists $C_0>0$ and $\epsilon >0$ such that for any $\xi_N$, 
\begin{eqnarray*}
P(\{x|K(x|\rho^X_{x,\xi_N},\xi_N,X)\leq p_X N\}|\xi_N,X)
+P(\{z|K(z|\rho^Z_{z,\xi_N},\xi_N,Z)\leq p_Z N\}|\xi_N,Z)
\leq 1+C_0 2^{-\epsilon N}
\end{eqnarray*}
holds for a sufficiently large $N$, 
where 
$P(\{x|K(x|\rho^X_{x,\xi_N},\xi_N,X)\leq p_X N\}|\xi_N,X)
:=\sum_x^{K(x|\rho^X_{x,\xi_N},\xi_N,X)\leq p_X N}P(x|\xi_N,X)$ and 
$P(\{z|K(z|\rho^Z_{z,\xi_N},\xi_N,Z)\leq p_Z N\}|\xi_N,Z)
:=\sum_z^{K(z|\rho^Z_{z,\xi_N},\xi_N,Z)\leq p_Z N}P(z|\xi_N,Z)$. 
\end{theorem}
The above theorem shows that there is a 
non-trivial trade-off relation between 
$P(\{x|K(x|\rho^X_{x,\xi_N},\xi_N,X)\leq p_X N\}|\xi_N,X)$ 
and $P(\{z|K(z|\rho^Z_{z,\xi_N},\xi_N,Z)\leq p_Z N\}|\xi_N,Z)$ 
when $q+p_X +p_Z<1$ holds. In particular, for sufficiently 
large $N$, if one of them is close to $1$, the other becomes 
exponentially small. 
In other words, as there is 
a pair $p_X$ and $p_Z$ satisfying $q+p_X +p_Z <1$ 
for any $q<1$, 
 there is a nontrivial effect of the 
bounded quantum memory for any $q<1$. 
\par
The above theorem is derived from 
the following lemma. 
\begin{lemma}\label{mainth}
Let us consider a protocol for a fixed $N$ and $M$. 
For any integers $l_X,l_Z \geq 0$, 
it holds: 
\begin{eqnarray*}
P(\{x|K(x|\rho^X_{x,\xi},\xi,X)\leq l_X\}|\xi,X)
+P(\{z|K(z|\rho^Z_{z,\xi},\xi,Z)\leq l_Z\}|\xi,Z)
\leq 1+2^{\frac{l_X +l_Z +M-N}{2}+c},
\end{eqnarray*}
where 
$P(\{x|K(x|\rho^X_{x,\xi},\xi,X)\leq l_X\}|\xi,X)
:=\sum_x^{K(x|\rho^X_{x,\xi},\xi,X)\leq l_X}P(x|\xi,X)$, 
$P(\{z|K(z|\rho^Z_{z,\xi},\xi,Z)\leq l_Z\}|\xi,Z)
:=\sum_z^{K(z|\rho^Z_{z,\xi},\xi,Z)\leq l_Z}P(z|\xi,Z)$ and 
$c$ is a constant depending on the choice of 
a quantum Turing machine. 
(Note that this inequality is non-trivial 
only for $\frac{l_X +l_Z +M-N}{2}+c <0$.)
\end{lemma}
{\bf Proof:}
\par
We
fix a universal quantum Turing machine $U$ and discuss 
the quantum Kolmogorov complexity with respect to it. 
We fix $\xi$ throughout the proof. 
Let us consider $K_U(z|\rho^Z_{z,\xi},\xi,Z)$ first. 
Thanks to theorem \ref{th:classical}, 
it suffices to consider only programs 
that exactly output the message $z$ because 
the message is a classical object. 
That is, we regard 
\begin{eqnarray*}
K_{c,U}(z|\rho_z,Z):=\min_{q: U(q,\rho^Z_{z,\xi},\xi,Z)=|z\rangle}
l(q),
\end{eqnarray*}
which satisfies $K_{c,U}(z|\rho^Z_{z,\xi},\xi,Z)\geq 
K_U(z|\rho^Z_{z,\xi},\xi,Z)
\geq K_{c,U}(z|\rho^Z_{z,\xi},\xi,Z)-c'$
for some constant $c'$. 
\par
Let us denote by $T^{\xi}_z \subset \{0,1\}^*$ a set of all programs 
that output $z$ with auxiliary inputs $\rho^Z_{z,\xi}$, $\xi$ and $Z$. 
That is, $T^{\xi}_z
=\{t \in \{0,1\}^* | 
U(t,\rho^Z_{z,\xi},\xi,Z)=z\}$ holds. 
An equation 
$K_{c,U}(z|\rho^Z_{z,\xi},\xi,Z)=
\min_{t\in T^{\xi}_z} l(t)$ follows.  
Although different programs may have different halting times, 
from the lemma 
proved by M\"uller (Lemma 2.3.4. in \cite{MullerPhD}),
we note that 
there exists a completely positive map (CP-map)
$\Gamma_{U,\xi,Z}: \Sigma({\cal H}_{m} \otimes {\cal H}_I)
\to \Sigma({\cal H}_O)$
satisfying for any $t\in T^{\xi}_z$ 
\begin{eqnarray*}
\Gamma_{U,\xi,Z}(\rho^Z_{z,\xi}
 \otimes |t \rangle \langle 
 t|)=|z\rangle \langle z|,
\end{eqnarray*}
where 
${\cal H}_I$ is 
a Hilbert space for 
programs, and ${\cal H}_O =\otimes^N {\bf C}_2$
is a Hilbert space 
for outputs.
From this theorem, we obtain an important 
observation. 
If $T^{\xi}_z \cap T^{\xi}_{z'} \neq \emptyset$ holds
for some $z\neq z'$, 
$\rho^Z_{z,\xi}$ and $\rho^Z_{z',\xi}$ are perfectly distinguishable.
In fact, because a CP-map does not increase 
the distinguishability of states, the relationships 
for $t\in T^{\xi}_z
\cap T^{\xi}_{z'}$
\begin{eqnarray*}
\Gamma_{U,\xi,Z}(\rho^Z_{z,\xi} \otimes |t\rangle \langle t|)
=|z\rangle \langle z| 
\\
\Gamma_{U,\xi,Z}(\rho^Z_{z',\xi} \otimes |t\rangle \langle t|)
=|z' \rangle \langle z'| 
\end{eqnarray*}
and their distinguishability on the right-hand sides 
imply the distinguishability of 
$\rho^Z_{z,\xi}$ and $\rho^Z_{z',\xi}$. 
For each $t\in \{0,1\}^*$, 
we define ${\cal C}^{\xi}_t \subset \{0,1\}^N$ 
as ${\cal C}^{\xi}_t=\{z| t\in T_z\}$. 
That is, $z\in {\cal C}^{\xi}_t$ is a message
that can be reconstructed by giving 
a program $t$ to the Turing machine $U$ with 
auxiliary inputs $\rho^Z_{z,\xi}$, $\xi$ and $Z$.  
Owing to the distinguishability between $\rho^Z_{z,\xi}$ 
and $\rho^Z_{z',\xi}$, for $z,z'\in {\cal C}^{\xi}_t$, 
there exists a family of projection operators 
$\{E^{t,\xi}_z\}_{z\in {\cal C}^{\xi}_t}$ on ${\cal H}_m$ satisfying
for any $z,z' \in {\cal C}^{\xi}_t$, 
\begin{eqnarray*}
&& E^{t,\xi}_z E^{t,\xi}_{z'} =\delta_{z z'} E^{t,\xi}_z
\\
&& \sum_{z\in {\cal C}^{\xi}_t}E^{t,\xi}_z \leq {\bf 1}
\\
&& \mbox{tr}(\rho_z E^{t,\xi}_{z'})=\delta_{zz'}.
\end{eqnarray*}
Because the memory is bounded as $\dim {\cal H}_m \leq 2^M$, 
\begin{eqnarray*}
\left| {\cal C}^{\xi}_t
\right| \leq 2^M
\end{eqnarray*}
holds. 
Because we are interested in minimum length 
programs, we define 
${\cal D}^{\xi}_t:=\{z| t=\mbox{argmin}_{s\in T^{\xi}_z} l(s)\}$, 
which 
is a set of all messages that  
have 
$t$ as the minimum length program for reconstruction. 
${\cal D}^{\xi}_t \subset {\cal C}^{\xi}_t$ holds. 
It is still possible that ${\cal D}^{\xi}_t 
\cap {\cal D}^{\xi}_{t'}
\neq \emptyset$. That is, there may be a message $z$ 
whose shortest 
programs are not unique. 
In such a case, we choose one of these programs to 
avoid counting doubly. 
For instance, this can be done by introducing 
a total order $<$ in all the programs $\{0,1\}^*$, 
and by defining 
${\cal E}^{\xi}_t=\{z|z\in {\cal D}^{\xi}_t, z \notin {\cal D}^{\xi}_{t'}
\mbox{ for all }t'<t\mbox{ with }l(t)=l(t')\}$. 
As this ${\cal E}^{\xi}_t$ is a subset of 
${\cal C}^{\xi}_t$, for any $z, z'\in {\cal E}^{\xi}_t$ 
\begin{eqnarray}
&& E^{t,\xi}_z E^{t,\xi}_{z'} =\delta_{z z'} E^{t,\xi}_z
\label{ortho} \\
&& \sum_{z\in {\cal E}^{\xi}_t}E^{t,\xi}_z \leq {\bf 1}
\label{pvmlike}
\\
&& \mbox{tr}(\rho_z E^{t,\xi}_{z'})=\delta_{zz'}
\nonumber
\end{eqnarray}
hold. 
The cardinality of ${\cal E}^{\xi}_t$ satisfies 
\begin{eqnarray}
\left|
{\cal E}^{\xi}_t 
\right|\leq 2^M.
\end{eqnarray}
\par
Similarly, we treat 
$K_U(x|\rho^X_{x,\xi},\xi,X)$. 
We can introduce 
$S^{\xi}_x \subset \{0,1\}^*$ as a set of all programs 
that output $x$ with auxiliary inputs $\rho^X_{x,\xi}$, $\xi$, and $X$. 
$K_{c,U}(x|\rho^X_{x,\xi},\xi,X)
=\min_{s\in S^{\xi}_x} l(s)$ holds. 
We can define ${\cal J}^{\xi}_s:=\{x|s \in S^{\xi}_x\}$ for each $s$
and introduce a family of 
projection operators $\{F^{s,\xi}_x\}_{x\in {\cal F}^{\xi}_s}$ 
on ${\cal H}_m$ 
that satisfies 
\begin{eqnarray*}
\mbox{tr}(F^{s,\xi}_x \rho^X_{x',\xi})=\delta_{xx'}
\end{eqnarray*} 
for each $x,x'\in {\cal J}^{\xi}_s$ and so on.
${\cal G}^{\xi}_s:=\{x| s=\mbox{argmin}_{t\in S^{\xi}_x} l(t) \}$
and 
${\cal F}^{\xi}_s=\{z|z\in {\cal G}^{\xi}_s, z \notin 
{\cal G}^{\xi}_{s'}
\mbox{ for all }s'<s\mbox{ with }l(s)=l(s')\}$, 
are also defined. 
An inequality for the cardinality, 
\begin{eqnarray}
\left|
{\cal F}^{\xi}_s
\right|
\leq \left|
{\cal J}^{\xi}_s
\right|
\leq 2^M
\label{Fcard}
\end{eqnarray}
also holds.
We consider a family of projection operators
$\{F^{s,\xi}_x \}_{x\in {\cal F}^{\xi}_s}$.
\par 
Let us analyze the protocol. 
Instead of the original protocol, 
we treat an entanglement-based protocol
(E91-like protocol \cite{E91}), 
which is equivalent to the original one.
It runs as follows. 
Alice prepares $N$ pairs of qubits. 
She prepares each pair in the EPR state, 
$|\phi\rangle:=
\frac{1}{\sqrt{2}}
(|0\rangle \otimes |0\rangle 
+|1\rangle \otimes |1\rangle)$. 
Therefore, the whole state can be 
written as $|\phi^N\rangle
:=|\phi\rangle \otimes |\phi \rangle 
\otimes \cdots \otimes |\phi \rangle$ 
($N$ times) 
in a Hilbert space ${\cal H}_{A'} \otimes {\cal H}_A
$, 
where ${\cal H}_{A'} \simeq {\cal H}_A \simeq \otimes^N {\bf C}^2$. 
Alice sends the qubits described by ${\cal H}_B$ to Bob. 
Bob makes the system ${\cal H}_A$ interact with 
his apparatus as
\begin{eqnarray*}
\Lambda: \Sigma({\cal H}_A) \to 
\Sigma({\cal H}_m \otimes {\cal K}).
\end{eqnarray*}
Bob measures his auxiliary system ${\cal K}$ with 
POVM $C=\{C_{\xi}\}$. 
When Bob has obtained $\xi$, we denote 
by $\Theta_{\xi}$ the a-posteriori state \cite{aposteriori}
over ${\cal H}_{A'} \otimes {\cal H}_m$. 
Alice measures her system ${\cal H}_{A'}$ with 
one of the bases $X$ or $Z$ to obtain a message. 
It can be shown that when Bob's obtained outcome was $\xi$ and 
Alice chose $X$ and obtained $x$, 
the state on the quantum memory ${\cal H}_m$ is 
$\rho^{X}_{x,\xi}$. 
Similarly, when Bob's obtained outcome was $\xi$ and 
Alice chose $Z$ and obtained $z$, 
the state on the quantum memory ${\cal H}_m$ is 
$\rho^{Z}_{z,\xi}$. 
\par
Hereafter we treat the state $\Theta_{\xi}$ over ${\cal H}_{A'}
\otimes {\cal H}_m$. 
Let us define a projection operator for each $t\in \{0,1\}^*$, 
$P^{\xi}_t :=\sum_{z\in {\cal E}^{\xi}_t}Z_z
\otimes E^{t,\xi}_z$, and for each $s \in \{0,1\}^*$, 
$Q^{\xi}_s :=\sum_{x \in {\cal F}^{\xi}_s}
X_x \otimes F^{s,\xi}_x$. 
For any integers $l'_X, l'_Z \geq 0$, 
we introduce projection operators, 
$\hat{P}^{\xi}_{l'_Z}:=\sum_{t}^{l(t)\leq l'_Z} P^{\xi}_t$ 
and $\hat{Q}^{\xi}_{l'_X}:=\sum_s^{l(s)\leq l'_X}
Q^{\xi}_s$. 
Their expectation 
values with respect to $\Theta_{\xi}$ can be written as:
\begin{eqnarray}
\mbox{tr}(\Theta_{\xi} \hat{P}^{\xi}_{l'_Z})
&=&P\left(
\{z| K_{c,U}(z|\rho^{Z}_{z,\xi},\xi,Z) \leq l'_Z\} 
| \xi,Z
\right)
\label{PlZ}
  \\
  \mbox{tr}(\Theta_{\xi} \hat{Q}^{\xi}_{l'_X})
&=&P\left(
\{x| K_{c,U}(x|\rho^{X}_{x,\xi},\xi,X) \leq l'_X\} 
| \xi,X
\right).
\label{QlX}
\end{eqnarray}
Our purpose is to find a trade-off inequality 
between (\ref{PlZ}) and (\ref{QlX}). It is obtained 
by the use of the uncertainty relation, which is 
one of the most important relation characterizing 
quantum mechanics. 
Among the various forms of uncertainty relations,
we employ the Landau-Pollak uncertainty relation 
\cite{MU,MiyaLP}. 
According to the generalized form given in 
\cite{MiyaLP}, for an arbitrary number of 
projections $\{A_i\}$, it holds that 
for any quantum state $\rho$, 
\begin{eqnarray*}
\sum_i \mbox{tr}(\rho A_i )
\leq 1 +
\left( \sum_{i\neq j}
\Vert A_i A_j\Vert^{2} 
\right)^{1/2}. 
\end{eqnarray*}
We apply this inequality for the state 
$\Theta_{\xi}$ and a family of positive operators 
$\{P_t,Q_s\}$ ($l(t)\leq l'_Z, l(s)\leq l'_X$). 
Because $P_t P_{t'} =Q_s Q_{s'}=0$ holds for $t\neq t'$ and 
$s\neq s'$ 
thanks to ${\cal E}^{\xi}_t \cap {\cal E}^{\xi}_{t'}
={\cal F}^{\xi}_s \cap {\cal F}^{\xi}_{s'}=\emptyset$, 
we obtain: 
\begin{eqnarray*}
\mbox{tr}(\Theta_{\xi}\hat{P}^{\xi}_{l'_Z})
+
\mbox{tr}(\Theta_{\xi} \hat{Q}^{\xi}_{l'_X})
\leq 1 +
\left(
2 \sum_t^{l(t)\leq l'_Z}\sum_s^{l(s)\leq l'_X}
\Vert 
Q^{\xi}_s P^{\xi}_t \Vert^2
\right)^{1/2}. 
\end{eqnarray*}
The term $\Vert Q^{\xi}_s P^{\xi}_t\Vert$ on the right-hand side 
is 
computed as follows. 
As the operator norm $\Vert Q^{\xi}_s P^{\xi}_t\Vert$ is 
written as $\Vert Q^{\xi}_s P^{\xi}_t \Vert 
=\sup_{|\Psi\rangle:\Vert \Psi\rangle \Vert=1}
\Vert Q^{\xi}_s P^{\xi}_t |\Psi\rangle \Vert$, 
we need to bound $\Vert  Q^{\xi}_s P^{\xi}_t
|\Psi\rangle \Vert$ for any normalized 
vector $|\Psi\rangle$. We consider:
\begin{eqnarray}
\Vert  Q^{\xi}_s P^{\xi}_t
|\Psi\rangle \Vert
&=&\langle \Psi |P^{\xi}_t Q^{\xi}_s P^{\xi}_t|\Psi\rangle^{1/2}. 
\label{pqpq}
\end{eqnarray}
As $Q^{\xi}_s =
\sum_{x \in {\cal F}^{\xi}_s}
X_x \otimes F^{s,\xi}_x
\leq \sum_{x \in {\cal F}^{\xi}_s}
X_x \otimes {\bf 1}_M$ holds, 
(\ref{pqpq}) can be bounded as
\begin{eqnarray*}
\langle \Psi |Q^{\xi}_s P^{\xi}_t Q^{\xi}_s|\Psi\rangle^{1/2}
&\leq & \left(
\sum_{z\in {\cal E}^{\xi}_t}
\sum_{x \in {\cal F}^{\xi}_s}
\sum_{z' \in {\cal E}^{\xi}_t}
\langle \Psi |
(Z_z
\otimes E^{t,\xi}_z )
( X_x \otimes {\bf 1}_M)
(Z_{z'}
\otimes E^{t,\xi}_{z'} )
|\Psi \rangle
\right)^{1/2}
\\
&= &
\left(
\sum_{z\in {\cal E}^{\xi}_t}
\sum_{x \in {\cal F}^{\xi}_s}
\langle \Psi |Z_z X_x Z_z 
\otimes E^{t,\xi}_z|\Psi \rangle 
\right)^{1/2}, 
\end{eqnarray*}
where we have used (\ref{ortho}). The right-hand side 
can be further deformed by introducing the a-posteriori state 
\cite{aposteriori}
$\mu^{t,\xi}_z$ as:
\begin{eqnarray*}
\left(
\sum_{z\in {\cal E}^{\xi}_t}
\sum_{x \in {\cal F}^{\xi}_s}
\langle \Psi |
Z_z X_x Z_z 
\otimes E^{t,\xi}_z|\Psi \rangle 
\right)^{1/2}
&=&
\left(
\sum_{z\in {\cal E}^{\xi}_t}
\sum_{x \in {\cal F}^{\xi}_s}
\mbox{tr}(\mu^{t,\xi}_z
Z_z X_x Z_z  )
\langle \Psi | {\bf 1}_B 
\otimes E^{t,\xi}_z|\Psi \rangle 
\right)^{1/2} 
\\
&\leq &
\left(
\sum_{z\in {\cal E}^{\xi}_t}
\sum_{x \in {\cal F}^{\xi}_s}
\frac{1}{2^N}
\langle \Psi | {\bf 1}_B 
\otimes E^{t,\xi}_z|\Psi \rangle 
\right)^{1/2} 
\\
&\leq& \left( \frac{1}{2^{N}}|{\cal F}^{\xi}_s|\right)^{1/2}
\leq 2^{\frac{M-N}{2}},
\end{eqnarray*}
where we have used a relation $|\mbox{tr}(\mu^{t,\xi}_z
Z_z X_x Z_z)| \leq \Vert Z_z X_x Z_z \Vert =\frac{1}{2^N}$, 
(\ref{pvmlike}) and (\ref{Fcard}). 
Thus we obtain $\Vert Q^{\xi}_s P^{\xi}_t\Vert
\leq 2^{\frac{M-N}{2}}$. 
Because $|\{t|l(t)\leq l'_Z\}|\leq 2^{l'_Z +1}$ and 
$|\{s|l(s)\leq l'_X\}|\leq 2^{l'_X +1}$ hold, 
we obtain 
\begin{eqnarray*}
\mbox{tr}(\Theta_{\xi}\hat{P}^{\xi}_{l'_Z})
+
\mbox{tr}(\Theta_{\xi} \hat{Q}^{\xi}_{l'_X})
\leq 1 +2^{\frac{l'_X +l'_Z +M -N +3}{2}}.
\end{eqnarray*} 
Taking into account the relation between 
$K_{c,U}$ and $K_U$ 
we finally obtain:
\begin{eqnarray*}
P(\{x|K(x|\rho^X_{x,\xi},\xi,X)\leq l_X\}|\xi,X)
+P(\{z|K(z|\rho^Z_{z,\xi},\xi,Z)\leq l_Z\}|\xi,Z)
\leq 1+2^{\frac{l_X +l_Z +M-N}{2}+c},
\end{eqnarray*}
where $c$ is a constant that depends on the choice of 
the quantum Turing machine. 
\hfill Q.E.D.
\par
{\bf Proof of theorem \ref{assym}:}
Apply lemma \ref{mainth} with $l_X=p_X N$, $l_Z=p_Z N$ and $M=q N$. 
As $c$ does not depend on $N$, we define $C_0:=2^{c}$. 
If we define $\epsilon:=\frac{1-(q+p_X+p_Z)}{2}$, 
we obtain the theorem. 
\hfill Q.E.D.
\par
In addition, the following corollary immediately follows: 
\begin{corollary}
Let us consider a protocol with fixed $N$. 
For any memory size $M\geq 0$ and any $\xi$, it holds
\begin{eqnarray*}
\max_{x:P(x|\xi,X)\neq 0} K(x|\rho^{X}_{x,\xi},\xi,X)
+ \max_{z:P(z|\xi,Z)\neq 0} K(z|\rho^{Z}_{z,\xi},\xi,Z)
\geq N-M+c.  
\end{eqnarray*}
\end{corollary}
{\bf Proof:}
From $P(\{x|K(x|\rho^X_{x,\xi},\xi,X)\leq 
\max_x K(x|\rho^{X}_{x,\xi},\xi,X) \}|\xi,X)=1$, 
$P(\{z|K(z|\rho^X_{z,\xi},\xi,Z)\leq 
\max_z K(z|\rho^{Z}_{z,\xi},\xi,Z) \}|\xi,Z)=1$, and 
the lemma \ref{mainth}, the claim immediately follows. 
(For $M=0$, it also holds.)
\hfill Q.E.D. 
\par
This corollary is regarded as a kind of 
generalizations of Heisenberg's uncertainty principle 
\cite{MiyaHeisen}. 
In fact, the case $M=0$ corresponds to the standard 
measurement scenario without a quantum memory. 
It implies that  
there is no observable that works as the joint measurement of 
$X$ and $Z$. 
\section{Discussion}
In this study, the power of the bounded quantum memory 
in simple information processing was examined. A 
toy two-party protocol was employed in which Bob, by using 
a bounded quantum memory and an unbounded classical memory, 
estimated 
messages encoded by Alice in one of the bases $X$ or $Z$. 
His ability to guess the message was characterized 
by the quantum Kolmogorov complexity. 
Our theorem provided a nontrivial effect of the memory boundedness. 
In addition, as a corollary, we obtained a generalization 
of the uncertainty principle with the quantum memory. 
As a future problem, it would be natural to apply the 
present study to the oblivious transfer. 
It should be noted that in \cite{Damgaard}, non-trivial 
result was obtained only for the quantum memory smaller than $N/2$ 
in the context of oblivious transfer, 
whereas our result indicates that there may be a non-trivial effect also 
for the quantum 
memory for which the size is $qN$ for some $q<1$. 
To investigate 
the complete protocol, we need to treat privacy amplification 
by using the Kolmogorov complexity. 
In addition, 
by combining the present result with the information-disturbance 
theorem \cite{MiyaEnt} in 
quantum key distribution, our theorem may play a role in estimating 
the threat of an eavesdropper who has a bounded quantum memory. 
We hope to investigate these problems in the future. 


\begin{thebibliography}{9}
\bibitem{prob}The word ``probabilistically" here is used to 
mean ``randomly in a probabilistic sense". 
That is, 
we use an unbiased probability $1/|\Omega|$ to choose a sample 
from a sample space $\Omega$ (say $\Omega=\{0,1\}^{2N}$).
(To avoid a 
possible confusion of it with randomness in 
algorithmic sense, we just write ``probabilistically".)
\bibitem{identify}
To treat $\rho^Z_{z,\xi}$ and $\rho^X_{x,\xi}$ as an auxiliary input
for a quantum Turing machine, they have to be somehow
represented as states on a system consisting of qubits. 
Our discussion does not depend on how we identify them.
\end{thebibliography}
\end{document}